\input harvmac\skip0=\baselineskip
\noblackbox

\def\bbbZ{\hbox{{Z$\!\!$Z}}}
\def\bR{\hbox{{I$\!$R}}}
\def\p{\partial}
\def\yt{\tilde{y}}

\def\tA{\tilde{A}}

\def\e{\epsilon}

\lref\msw{ J.~M.~Maldacena, A.~Strominger, E.~Witten,
  ``Black hole entropy in M-theory''
  JHEP {\bf 9712}, 002 (1997)
  [arXiv:hep-th/9711053].

}
\lref\denef{F. Denef, Talk given at Harvard Workshop on Black Holes and Topological Strings, Jan. 31 '06.}

  \lref\gravmul{
  A.~Fujii, R.~Kemmoku and S.~Mizoguchi,
  ``D = 5 simple supergravity on AdS(3) x S(2) and N = 4 superconformal  field
  theory,''
  Nucl.\ Phys.\ B {\bf 574}, 691 (2000)
  [arXiv:hep-th/9811147].
  %%CITATION = HEP-TH 9811147;%%
  %%Cited 11 time in SPIRES-HEP
}

\lref\ascv{A.~Strominger, C.~ Vafa,  ``Microscopic origin of the Bekenstein-Hawking entropy''
 Phys.Lett. B {\bf  379} (1996),  99-104 
[arXiv: hep-th/9601029]. }

\lref\bmpv{J.C.~Breckenridge, R.C. Myers, A.W. Peet, C. Vafa, ``D-branes and spinning black holes''
Phys.Lett.B {\bf 391} (1997), 93-98 [arXiv: hep-th/9602065].}

\lref\cv{C. Vafa, ``Black holes and Calabi-Yau threefolds''
  Adv.Theor.Math.Phys. {\bf 2} (1998) 207-218 
[arXiv: hep-th/9711067].}

\lref\jmas{J.M. Maldacena, A. Strominger, ``AdS(3) black holes and a stringy exclusion principle''
 JHEP {\bf 9812} (1998) 005   [arXiv: hep-th/9804085].}

\lref\fourdfived{D. Gaiotto, A. Strominger, X. Yin, ``New connections between 4-D and 5-D black holes'' JHEP {\bf 0602} (2006) 024 [arXiv: hep-th/0503217].}

\lref\ortin{N.~Alonso-Alberca, E. Lozano-Tellechea, T. Ortin, ``The near horizon limit of the extreme rotating D = 5 black hole as a homogeneous space-time'' Class.Quant.Grav. {\bf 20} (2003) 423-430
 [arXiv: hep-th/0209069].}
\lref\gibb{G.W. Gibbons, C.A.R. Herdeiro, ``Supersymmetric rotating black holes and causality violation'' Class.Quant.Grav. {\bf16} (1999) 3619-3652 
[arXiv: hep-th/9906098].}
\lref\ldys{L. Dyson, ``Studies Of The Over-Rotating BMPV Solution'' [arXiv: hep-th/0608137].}

\lref\fairy{J. de Boer, M.C.N. Cheng, R. Dijkgraaf, J. Manschot, E. Verlinde, ``A Farey Tail for Attractor Black Holes'' JHEP {\bf 0611} (2006) 024 
[arXiv: hep-th/0608059], and talk given by E. Verlinde at Strings '06.}

\lref\larsen{M. Cvetic, F. Larsen, ``Near horizon geometry of rotating black holes in five-dimensions'' Nucl.Phys.B {\bf 531} (1998) 239-255 
[arXiv: hep-th/9805097].}

\lref\liwei{W. Li, A. Strominger, ``Supersymmetric probes in a rotating 5D attractor''
[arXiv: hep-th/0605139].}

%\draft

\Title{\vbox{\baselineskip12pt\hbox{} }} {Wrapped M2/M5 Duality}  \centerline{Monica Guica and Andrew Strominger}
\smallskip
\centerline{Jefferson Physical Laboratory, Harvard University,
Cambridge, MA 02138} \vskip .6in \centerline{\bf Abstract} {A microscopic accounting  of the entropy of a generic 5D supersymmetric rotating black hole, arising from wrapped $M2$-branes in Calabi-Yau compactified $M$-theory,  is an outstanding unsolved problem. In this paper we consider an expansion around the zero-entropy, zero-temperature, maximally rotating ground state for which the angular momentum $J_L$ and graviphoton charge $Q$ are related by $J_L^2=Q^3$. At $J_L=0$ the near horizon  geometry is $AdS_2\times S^3$. As $J_L^2 \to Q^3$ it becomes a singular quotient of $AdS_3\times S^2$: more precisely, a quotient of 
the near horizon geometry of an $M5$ wrapped on a 4-cycle whose self-intersection is the 2-cycle associated to the wrapped-$M2$ black hole. The 
singularity of the $AdS_3$ quotient is identified as the usual one associated to the zero-temperature limit, suggesting that the $(0,4)$ wrapped-$M5$ CFT is  dual
near maximality to the wrapped-$M2$ black hole.  As evidence for this, the microscopic  $(0,4)$ CFT entropy  
and the  macroscopic rotating black hole entropy are 
found to agree to leading order away from maximality. 

 } \vskip .3in

%\smallskip
\Date{}\listtoc \writetoc
\newsec{Introduction}
    The microscopic origin of the entropy of rotating 5D BPS black holes - known as BMPV black holes - in 
    $N=8$ and $N=4$ theories has been understood for some time \refs{\ascv, \bmpv}.  In contrast,  the origin of the entropy in 5D $N=2$ theories, in particular  those arising from wrapped $M2$-branes in Calabi-Yau compactified $M$-theory, has remained enigmatic (except in some special cases \cv).   In four dimensions, as long as the $D6$ charge $p^0$ vanishes, the entropy has been understood in terms of the 
 microstates of a certain $(0,4)$ CFT - known as the MSW CFT \msw\  - that arises from a wrapped $M5$-brane.  However, when $p^0\neq 0$, the problem again remains unsolved. 
Due to  the 4D-5D connection \fourdfived, the unsolved 5D problem is equivalent to the unsolved 4D case with $p^0=1$. 

  Naively, one might expect  the $SU(2)_L$ angular momentum $J_L$ to  be carried by excitations of the
  5D black hole as in the $N=4,8$ cases\foot{In these cases the angular momentum $J_L$ is carried by left moving fermions of the $D1-D5$ CFT.}  \refs{\ascv, \bmpv} and try to understand the $J_L=0$ ground state by finding a dual of 
  the near horizon $AdS_2\times S^3$. However the
  area-entropy law
  \eqn\rti{S_{BH}=2 \pi \sqrt{Q^3-J_L^2}}
  where $Q$ is the graviphoton charge, suggests that the ground state 
of the system is at maximality, where $J_L^2=Q^3$. The behaviour of the near maximal-solutions has been studied in \refs{\gibb\ortin-\ldys}. Despite the fact that the area goes to zero, the geometry remains smooth, while a $U(1)_L$ orbit of the horizon becomes null.  Locally, the geometry becomes that of $AdS_3\times S^2$ \ortin, with curvatures and fluxes corresponding to an $M5$-brane wrapped on a particular 4-cycle in the Calabi-Yau. The particular 4-cycle is the one whose self-intersection is the 2-cycle wrapped by the $M2$-brane carrying the black hole charges \foot{Not every 2-cycle is the self-intersection of an integral 4-cycle. In the following we restrict our attention to those 2-cycles for which this is the case.  A dual relation - probably related to the one discussed herein -  between such $M2$ and $M5$ branes  was discussed in \fairy.}. 
     Globally, the near-maximal BMPV is a quotient of $AdS_3\times S^2$. At maximality, the orbits of the quotient have zero proper length, which looks a bit singular.  However, exactly this type of singularity has appeared before \jmas. 
The quotient of $AdS_3$ by the $SL(2, \bR )_L$ element which takes $w^+ \to 
e^{4\pi^2 T_L}w^+$ has zero length orbits as $T_L \to 0$.  This quotient is known to be dual to a mixed state of the  boundary CFT in which the left movers are at temperature $T_L$.

   This all suggests that  the dual of the BMPV black hole is related to the MSW CFT which sits at the boundary of the limiting $AdS_3\times S^2$, with a temperature for left-movers determined by the deviation from maximality via the limiting form of the quotient.   We will show in section 2  that this indeed produces the correct expression \rti \  for the entropy to leading order away from maximality.

Expanding the geometry to second order away from maximality, we find a 
deformation of $AdS_3\times S^2$ in which the $U(1)$ fiber associated to the graviphoton mixes with the spacetime directions. We have not succeeded in interpreting this, but it suggests that  the exact dual theory beyond leading order away from maximality is some kind of deformation of the MSW CFT.  An understanding of the general case would be of great interest. 

In section 3 we consider some $Z_{p^0}$ quotients of BMPV which arise as the 5D lift of $D6$ charged 4D black holes.  For special values of the charges
we find that these can be mapped to quotients of $AdS_3\times S^2$ even far from maximality, and that the microscopic entropy arising from the associated MSW CFT agrees with the Bekenstein-Hawking result. 

Some relevant $AdS_3$ coordinate transformations are recorded in an appendix.

\newsec{Near-maximal entropy}

\subsec{BMPV$ \to AdS_3\times S^2$}

In this section we show that the total space of the graviphoton 
fiber bundle over BMPV is locally the same as the total space of the graviphoton fiber bundle over $AdS_3\times S^2$ for a certain relation between the graviphoton fluxes,  and that both are globally quotients of   $AdS_3\times S^3$. We first consider the case of a single $U(1)$ gauge field and then generalize to $N$ gauge fields. 

The metric for the near-horizon BMPV is\foot{  The $t$ coordinate used here differs by a factor of $\cos B$ from the usual asymptotic time coordinate. We set
$\ell_5=({4G_5 \over \pi})^{1/3}=1$.} (in Poincar\'e  coordinates)
\eqn\pmet{ ds_5^2={Q \over 4}\left[-(\cos B{ dt\over
\sigma}+\sin{B}\sigma_3)^2+ {d\sigma^2 \over
\sigma^2}+\sigma^2_1+\sigma^2_2+ \sigma^2_3 \right]}where the
right-invariant one-forms are \eqn\scb{\eqalign{
\sigma_1&=-\sin{\psi}d\theta+\cos{\psi}\sin{\theta}d\phi\cr
\sigma_2&=\cos{\psi}d\theta+\sin{\psi}\sin{\theta}d\phi\cr
\sigma_3&=d\psi+\cos{\theta}d\phi}} $\psi$ is identified modulo
$4 \pi $ and  the parameter  $B$ is related to the angular momentum  $J_L$ by 
\eqn\avo{\sin B = {J_L \over Q^{3/2}}}  At maximal rotation, $\sin^2 B=1$ and 
$J_L^2=Q^3$. The graviphoton field strength in the coordinates \pmet\ is\foot{Our normalization is such that ${1 \over 2 \pi }\int F$ is integral. }
\eqn\grv{ F=dA, \qquad \qquad A={{{\sqrt{Q}}}\over 2}(\cos B{dt \over
\sigma}+\sin{B}\sigma_3)}

\noindent{ $\bullet$ \bf One $\bf U(1)$}

The total space of the $U(1)$ bundle over
BMPV defines a six manifold with metric \eqn\pemet{\eqalign{
ds_6^2&={Q \over 4}\left[-(\cos B{dt\over \sigma}+\sin{B}\sigma_3)^2+
{d\sigma^2 \over
\sigma^2}+\sigma^2_1+\sigma^2_2+\sigma^2_3\right]+(dy+A)^2\cr &={Q
\over 4}\left[ {4 \over Q}dy^2 +{ 4 \over {\sqrt{Q}}}dy(\cos B{dt\over
\sigma}+\sin{B}\sigma_3)+ {d\sigma^2 \over
\sigma^2}+\sigma^2_1+\sigma^2_2+\sigma^2_3\right]}} 

where $y \sim
y+2\pi n$ is the fiber coordinate. Defining \eqn\rty{\eqalign{x &=  { 2 \cos B \over {\sqrt{Q}}} y\cr
    \yt & ={\sqrt{Q} \over 2} \psi +\sin B y}}

one finds \eqn\mlk{ds_6^2={Q \over 4}\bigl(dx^2+{2 d x
d  t\over \sigma}+ {d\sigma^2 \over
\sigma^2}+\sigma^2_1+\sigma^2_2 \bigr) +(d\tilde y+\tilde A)^2} 
with
\eqn\dsv{\tilde A ={\sqrt{Q} \over 2}\sin \theta d\phi}
This is precisely the $AdS_3 \times S^2$ near horizon metric for the charge 
$p=\sqrt Q$ MSW string, provided $p$ is an integer. Hence we see that a coordinate transformation which mixes up the base and fiber transforms the near horizon geometry of $p^2$ wrapped $M2$-branes to that of $p$ wrapped $M5$-branes.  

\bigskip
\noindent{$\bullet$ \bf $\bf N$  $\bf U(1)$s}

Now let us repeat the discussion for the case of $N-1$ vector
multiplets and associated black hole charges $q_A$.  We define $p^A$
as the 4-cycle whose self-intersection is $q_A$,
that is \eqn\rsz{q_A=3D_{ABC}p^Bp^C } 
The graviphoton charge is then
\eqn\dsc{Q^{3/2}=D_{ABC}p^Ap^Bp^C} The attractor values of the
vector moduli are $t^A=p^A/\sqrt{Q}$. We assume that  $q_A$ are such
that the $p^A$ are integers. (This is not the case for all integral
$q_A$). The metric is still given by \pmet, but now the gauge fields
are \eqn\dli{F^A=dA^A, \qquad \qquad A^A={p^A\over 2}(\cos B {d
t \over \sigma}+\sin{B}\,\sigma_3)} The total space of the $U(1)^N$
bundle over BMPV has the  metric \eqn\pmest{\eqalign{ ds_6^2&={Q \over
4} \left[-(\cos B{dt\over \sigma}+\sin{B}\sigma_3)^2 + {d\sigma^2 \over\sigma^2}+\sigma^2_1+\sigma^2_2+\sigma^2_3\right]\cr &~~~~~~~~~~~~~~~~+(2t_At_B -
D_{AB})(dy^A+A^A)(dy^B+A^B)} }where \eqn\edf{D_{AB}\equiv
D_{ABC}t^C~,~~~~ t_A \equiv D_{AB}t^B~,~~~~y^A\sim y^A+2\pi
n^A,~~n^A \in \bbbZ } Defining \eqn\fsx{\tilde y^A \equiv y^A+(\sin B-1) t^A
t_B y^B + \half p^A \psi}
\eqn\jjh{x={2 \cos B \over \sqrt{Q}} t_Ay^A}we can write the metric as
\eqn\pmst{\eqalign{ ds_6^2&={Q \over 4}(dx^2 +{2dx d t\over \sigma}+ {d\sigma^2 \over
\sigma^2}+\sigma^2_1+\sigma^2_2 )+(2t_At_B - D_{AB})(d\tilde
y^A+\tA^A)(d\tilde y^B+\tA^B) }} with \eqn\osa{\tA^A=\half \,p^A
\cos \theta d\phi } This is an N-dimensional fibre bundle over $AdS_3
\times S^2$,  with the following identifications \eqn\esd{\eqalign{
\tilde y^A& \sim \tilde y^A+2\pi n^A+(\sin B -1)2\pi
t^At_{C}n^C+2\pi m p^A \cr x &\sim x +{4 \pi\cos B
\over \sqrt{Q}}t_A n^A }} 

\noindent{$\bf AdS_3\times S^3$}

We have just seen that the total bundle space of near horizon wrapped $M2$'s and wrapped  $M5$'s  
are equivalent. We now see  that they are both quotients of 
$AdS_3\times S^3$  with a flat $U(1)^{N-1}$ bundle\foot{Closely related observations were made in \larsen.}. 
Defining 
\eqn\rao{\eqalign{\tilde \psi& = \psi+{2\sin B  \over \sqrt Q} t_A y^A \cr~~~~~~\tilde \sigma_3
&=d\tilde \psi+\cos \theta d\phi \cr y^A_L&=y^A-t^At_By^B}}
and substituting in the BMPV metric with $U(1)^N$  bundle \pmest\ we get   \eqn\mldk{ds_6^2={Q \over 4}\bigl(dx^2+{2 d x
d  t\over \sigma}+ {d\sigma^2 \over
\sigma^2}+\sigma^2_1+\sigma^2_2 +\tilde \sigma_3^2 \bigr)-D_{AB} dy^A_L dy^B_L } 
with the identifications \eqn\esd{\eqalign{\tilde
\psi &\sim \tilde \psi+ 4\pi {\sin B\over \sqrt Q}t_An^A + 4 \pi m \cr
x  &\sim x+4 \pi {\cos B \over \sqrt Q} t_An^A \cr y_L^A & \sim y_L^A + 2 \pi n^A - 2 \pi t^A t_B n^B}}
Of the six dimensions in \mldk, five are geometrical and the sixth is 
the $U(1)$ fiber. The fiber $U(1)$ is generated by the unit-normalized Killing vector 
\eqn\wws{K_F=t^A{\p\over \p y^A} ={2 \cos B  \over \sqrt Q} \p_x+{2 \sin B  \over \sqrt Q} \p_{\tilde \psi}}

\subsec{Near-maximal BMPV entropy}

To get from $AdS_3\times S^3$ in \mldk\ to $AdS_3\times S^2$, 
we regard the $S^3$  factor as a $U(1)$ bundle over $S^2$ and reduce along 
$\tilde \psi$. 
To get to nonrotating BMPV, we regard the $AdS_3$ factor as a 
$U(1)$ bundle over $AdS_2$ and reduce along $x$.  For rotating BMPV, we reduce along the linear combination of $x$ and $\tilde \psi$ determined by the Killing vector \wws.   At maximal rotation, $\sin B=1$ and $\cos B=0$, so the reduction is entirely along $\tilde \psi$. This demonstrates that maximal BMPV is locally $AdS_3\times S^2$. 

However, the relation between the generic BMPV and $AdS_3\times S^2$  can not be so 
direct because they have different topologies. What happens is that the quotient 
\esd\  becomes singular in the maximal limit $J_L^2\to Q^3$, $\cos B \to 0$. 
This turns out to  be exactly what is needed: the singularity of the quotient is precisely of the form expected from the fact that the Hawking temperature $T_L$ (for the left movers of the MSW CFT) goes to zero at maximality.

To see this we write  $B ={\pi \over 2}-\epsilon$ and define $n^A = \hat{n}^A -
  m p^A$.  Keeping only the leading terms in $\e$, \esd\ reduces to
\eqn\rop{\eqalign{\tilde y^A& \sim \tilde y^A+2\pi \hat{n}^A \cr x & \sim x - 4 \pi \e \,m}} Writing the identification this way
it becomes clear that in the maximally-rotating limit the
$\hat{n}^A$ identification only acts on the $N$ fiber coordinates, while
the $m$ identification only acts on $AdS_3$.
Moreover, the $AdS_3$ identification lives purely
in the $SL(2,\bR)_L$ factor of the isometry group. 
$AdS_3$ quotients of this type were analyzed in \jmas. While $t$ is a standard null coordinate on the Poincar\'e diamond of 
the boundary of $AdS_3$, $x$ in fact turns out to be a Rindler coordinate.  The $SL(2,\bR )_L$-invariant $AdS_3$ vacuum then gives,
after $x $ identifications,  a thermal state for the left movers of the boundary CFT ( in this case the MSW CFT).  This quotient produces a boundary CFT in a thermal ensemble
with right moving temperature $T_R=0$ and left moving temperature
equal to  the magnitude of the $x$ shift divided by $4\pi^2$. Hence from \rop\ we have  
\eqn\ghy{T_L={ \epsilon \over \pi }} As expected, $T_L\to 0 $ at maximality and the quotient becomes singular.
The entropy of the MSW string at temperature $T_L$ is
\eqn\ykl{S_{micro}={\pi^2 c_LT_L\over 3}={2\pi D_{ABC}p^Ap^Bp^C \epsilon }=2\pi
Q^{3/2}\epsilon} The Bekenstein-Hawking entropy of 
BMPV is \eqn\dwa{S_{macro}=2 \pi Q^{3/2}\cos B } which agrees with
\ykl\ near maximality.  
  This supports the conjecture that the excitations of the BMPV black hole near its maximally spinning ground state are dual to 
to the MSW string on the boundary of the limiting $AdS_3\times S^2\times CY$ attractor. 

\subsec{Far from maximality}

In the above derivation, we made crucial use of the fact that the
black hole was near maximal rotation 
and that the discrete identifications could be approximated 
by the leading terms in the near-maximal expansion parameterized by 
$\epsilon= {\pi \over 2}-B$.  In this approximation the resulting geometry has a simple interpretation as a well-studied quotient of 
$AdS_3$. However at next to leading order, the identifications mix up the base and fiber coordinates in a way that makes the spacetime interpretation obscure. It is quite possible that there is some way of making sense of the general case, but we have not succeeded in doing so. This indicates that, while the MSW string provides a good description of near-maximal excitations, it is not the exact dual of BMPV. Perhaps the dual is some kind of deformation of MSW. 

It turns out that one can do a bit better - for the case of a single $U(1)$ - by considering quotients of BMPV. This will be the subject of the next section.

\newsec{General D6 charge with a single $U(1)$}

In the preceding section we argued that the BPS excitations of a BMPV black hole very near its maximally rotating ground state were described by the dual MSW string.  Using the 4D-5D connection \fourdfived\
these become excitations of the maximally $D0$-charged 4D black hole with $D6$ charge $p^0=1$.  In this section we will consider general $p^0$, but a single $U(1)$. We will find -for certain charges -
evidence for a general relation to the MSW string not limited to near maximality.  We find it clearest to start with the near horizon $M5$ geometry known to be dual to MSW, and show that it can be recast 
as a  quotient of BMPV.

We start with a well-understood quotient of $AdS_3\times S^2$ (equation \mlk) accompanied by a twist in the graviphoton connection, which can generically be written as
\eqn\tws{\eqalign{x & \sim x + 4 \pi^2 T_L k \cr \yt  & \sim
\yt + 2 \pi w + 2 \pi k \theta }} $\theta$ is a chemical potential for graviphoton charge. We then use the bundle coordinate transformations \fsx,\jjh\ to local BMPV bundle coordinates and ask for which values of
$T_L$ and $\theta$ this system actually
describes BMPV.  Of course, bundle coordinate transformations 
which mix up the base and the fiber are not known in general to be a symmetry of string theory (unless of course the $U(1)$ is a Kaluza-Klein $U(1)$), so it is not guaranteed that this will give a sensible result. Hence our final result, which does appear sensible, must be regarded as a conjecture.\foot{We note that in the near maximal discussion of the previous section, the coordinate transformations did not, at leading order, mix the fiber and base. }  
The identifications \tws\ in BMPV coordinates become
\eqn\ifc{\eqalign{y & \sim y + {2
\pi^2 {p} T_L k \over \cos B}  \cr \psi & \sim \psi + {4 \pi w
\over {p}} + {4 \pi k }({\theta \over {p}} -  \pi T_L 
\tan B)}} where $p^2=Q$. In order for the graviphoton charge quantization condition to be the one dictated by $M$ theory, the $y$ fiber must be identified modulo $2 \pi  k$, which  requires 
\eqn\oook{T_L = {\cos B
\over \pi {p}}}
In order for $y$ to be a true fiber over BMPV, the $y$ identifications generated by $k$ must not involve a shift in the base. This requires 
\eqn\rko{\theta=\sin B .}
We are then left with the identifications
 \eqn\tht{y \sim y+2\pi k,~~~~~
 \psi \sim \psi  + {4 \pi w
\over {p}}  } 
This contrasts with usual BMPV for which one has $\psi \sim \psi+4\pi w$. However \tht\  has a simple interpretation.  The factor of $p$ in \tht\ means that we have BMPV with an $S^3/Z_p$ horizon.  

The $Z_p$ quotient is the near horizon geometry of the 5D lift of 
a 4D black hole with $D6$ charge $p^0=p$ and $q_0=2J_L$.  This has Bekenstein Hawking entropy  
\eqn \saxc{S_{BH}= 2 \pi {Q^{3/2}\cos B \over p^0}= 2 \pi p^2 \cos B}
On the other hand this was related to the MSW string at temperature 
\oook. This has an entropy 
\eqn\daij{S_{micro}= {\pi^2 \over 3} c_L T_L =
2 \pi p^2 \cos B=S_{BH}}
This motivates the conjectured duality between 
the 4D black hole with charges $(p^0,p^1,q_1,q_0)=(p,0,p, q_0)$ 
and the MSW CFT of charge $p$ $M5$-branes at temperature 
\eqn\tleft{T_L= {1\over \pi p} \sqrt{1- {q_0^2 \over 4 p^2}}}.

\centerline{ \bf Acknowledgements}

We are grateful to Davide Gaiotto, Lisa Huang, Wei Li, Erik Verlinde and Xi Yin for helpful discussions. This work has been supported in part by the DOE grant DE-FG02-91ER40654.

\appendix{A}{}
We list herein the transformations from the $AdS_3$ coordinates in \mlk\ to the Poincar\'e coordinates used in \jmas. In Poincar\'e coordinates, the metric reads

\eqn\poim{ds^2=Q {dw^+dw^-+d\xi^2 \over \xi^2} + {Q \over 4} (\sigma_1^2
+ \sigma_2^2 + \tilde{\sigma}_3^2)} The transformation from
\mlk\ to Poincar\'e is

 \eqn\kis{\eqalign{w_+ &= e^{x}\cr w_- &= \half({t} - \sigma)\cr \xi^2 &= \sigma e^{x}}}

\listrefs

\end